\documentclass[12pt]{article}
\usepackage{amsmath,amssymb,graphicx,array}
\usepackage{alltt,url,thophys,emp}
\empprelude{input boxes}
\empprelude{input rboxes}
  {\bgroup\footnotesize\alltt}%
  {\endalltt\egroup\noindent}
\begin{document}
\begin{empfile}
\title{%
  \parbox{\textwidth}{%
    \begin{flushright}\texttt{IKDA 2000/31}\end{flushright}}\\[\baselineskip]
  O'Mega~\&~WHIZARD:
  Monte~Carlo Event~Generator~Generation For~Future~Colliders}
\author{Thorsten Ohl\\\hfil\\
  Darmstadt University of Technology\\
  Schlo\ss{}gartenstra\ss{}e 9, 64289 Darmstadt, Germany\\
  \texttt{ohl@hep.tu-darmstadt.de}}
\maketitle
\begin{abstract}
  I describe the optimizing matrix element generator
  \textit{O'Mega}~\cite{Ohl:2000:ACAT,O'Mega} and  Wolfgang Kilian's
  event generator generator \textit{WHIZARD}~\cite{Kilian:WHIZARD}.
  These tools cooperate in the automated production of efficient
  unweighted event generators for linear collider physics.
\end{abstract}

\section{Introduction}

Current and planned experiments in high energy physics can probe
processes with many tagged---potentially polarized---particles in the
final state, in particular at a linear collider.  The combinatorial
explosion of the number of Feynman diagrams contributing to scattering
amplitudes for many external particles calls for the development of
more compact representations that translate well to efficient and
reliable numerical code.  In gauge theories, strong numerical
cancellations in a redundant representation built from necessarily
gauge dependent Feynman diagrams lead to a loss of numerical
precision, stressing further the need for eliminating redundancies.

At the same time, the final state phase space becomes more and more
intricate and efficient Monte Carlo sampling turns into a highly
non-trivial problem.  The rapidly growing number of nonfactorizable
singularities poses a challenge to adaptive sampling algorithms
(see~\cite{VAMP}).

Due to the large number of processes that have to be studied in order
to unleash the potential of current and planned experiments, including
a linear collider, the construction of optimized representations of
scattering amplitudes must be possible algorithmically on a computer
and should not require human ingenuity for each new application.  For
the same reason, improved phase space sampling algorithms should be
adaptive and robust, allowing the construction of unweighted event
generators with a minimum of human intervention.

\section{O'Mega}
O'Mega~\cite{Ohl:2000:ACAT,O'Mega} is a generator for tree-level
scattering amplitudes that satisfies the requirements set forth in the
introduction.  O'Mega constructs the scattering amplitude from
\emph{One Particle Off-shell Wave functions}~(1POWs)
\begin{equation}
  W_{p_1,\ldots,p_n}^{q_1,\ldots,q_m}(x) =
     \Braket{\phi(q_1),\ldots,\phi(q_m);out|\Phi(x)
              |\phi(p_1),\ldots,\phi(p_n);in}\,.
\end{equation}
The 1POWs are sums of Feynman diagrams. Therefore, expressing the
scattering amplitude in terms of the 1POWs achieves a factorization of
the sum of all Feynman diagrams.  Indeed, this representation removes
all redundancies and dramatically reduces the growth in calculational
effort from a factorial of the number of particles to an exponential.
O'Mega can emulate both numerical approaches
in~\cite{ALPHA:1997/HELAC:2000} and produces code that is empirically
at least twice as fast.  In addition, the symbolic nature of O'Mega
provides greater flexibility in the translation to numerical code:
treatment of unstable particles, verification of Ward identities,
etc.

O'Mega is independent of the target language and can support code in
any programming language for which a simple output module has been
written.  The code generated by the Fortran90/95 backend is the most
efficient code available for polarized scattering amplitudes for many
particles.  To support a physics model, O'Mega requires as input only
the Feynman rules and the relations among coupling constants.
Currently, the standard model is well tested (the numerical results
agree with MADGRAPH~\cite{MADGRAPH:1994}) and the MSSM is in
preparation.

\section{WHIZARD}
\begin{empcmds}
  def vconnect (suffix from, ff, to, tf) =
    (ff[from.sw,from.se]{down}---{down}tf[to.nw,to.ne]
     cutbefore bpath.from cutafter bpath.to)
  enddef;
  def hconnect (suffix from, ff, to, tf) =
    (ff[from.ne,from.se]{right}---{right}tf[to.nw,to.sw]
     cutbefore bpath.from cutafter bpath.to)
  enddef;
  def fixsizepos (text t) =
    fixsize (t);
    fixpos (t)
  enddef;
\end{empcmds}
\begin{figure}
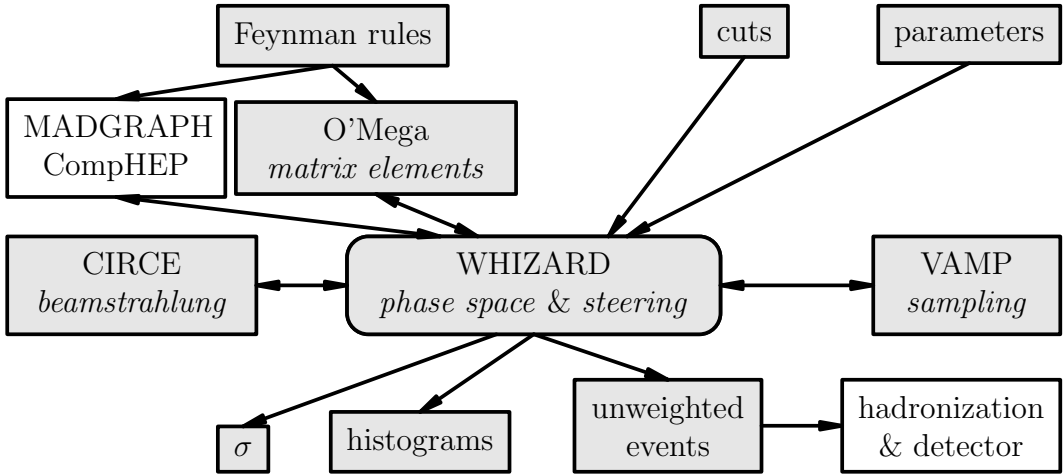

  \begin{center}
    \begin{empx}(140,62)
      \empcmd{ahlength := 3mm;}
      \empcmd{ahangle := 20;}
      \empcmd{pickup pencircle scaled 1.5pt;}
      \empcmd{defaultdx := 6pt;}
      \empcmd{defaultdy := 6pt;}
      \empcmd{boxit.feynman (btex Feynman rules etex);}
      \empcmd{boxit.params  (btex parameters etex);}
      \empcmd{boxit.cuts  (btex cuts etex);}
      \empcmd{boxit.omega (btex \parbox{8em}{\centering
                                  O'Mega\\ \emph{matrix elements}} etex);}
      \empcmd{boxit.madgraph (btex \parbox{6em}{\centering
                                     MADGRAPH\\ CompHEP} etex);}
      \empcmd{rboxit.whizard (btex \parbox{11em}{\centering
                                      WHIZARD\\ \emph{phase space} \&\ \emph{steering}} etex);}
      \empcmd{boxit.vamp (btex \parbox{5em}{\centering VAMP\\ \emph{sampling}} etex);}
      \empcmd{boxit.circe (btex \parbox{7em}{\centering CIRCE\\ \emph{beamstrahlung}} etex);}
      \empcmd{boxit.sigma (btex $\sigma$ etex);}
      \empcmd{boxit.hist (btex histograms etex);}
      \empcmd{boxit.events (btex \parbox{5em}{\centering unweighted\\ events} etex);}
      \empcmd{boxit.jetset (btex \parbox{6em}{\centering hadronization\\ \&\ detector} etex);}
      \empcmd{feynman.nw = (.2w,h);}
      \empcmd{params.ne = (w,h);}
      \empcmd{cuts.n = (.7w,h);}
      \empcmd{madgraph.w = (0, whatever);}
      \empcmd{madgraph.c = .55[whizard.c,feynman.c] + (whatever,0);}
      \empcmd{madgraph.e = omega.w - (.01w,0);}
      \empcmd{circe.w = (0,.4h);}
      \empcmd{whizard.c = (.5w,.4h);}
      \empcmd{vamp.e = (w,.4h);}
      \empcmd{sigma.sw = (.2w,0);}
      \empcmd{hist.s = .25[sigma.s,jetset.s];}
      \empcmd{events.s = .6[sigma.s,jetset.s];}
      \empcmd{jetset.se = (w,0);}
      \empcmd{fixsizepos (whizard, feynman, params, cuts, sigma, hist,
                          events, jetset, vamp, omega, madgraph);}
        {\empcmd{fill bpath whizard withcolor 0.9white;}%
         \empcmd{drawboxed (whizard);}}
        {\empcmd{fill bpath feynman withcolor 0.9white;}%
         \empcmd{fill bpath params withcolor 0.9white;}%
         \empcmd{fill bpath cuts withcolor 0.9white;}%
         \empcmd{drawboxed (feynman, params, cuts);}}
        {\empcmd{fill bpath sigma withcolor 0.9white;}%
         \empcmd{fill bpath hist withcolor 0.9white;}%
         \empcmd{fill bpath events withcolor 0.9white;}%
         \empcmd{drawboxed (sigma, hist, events);}}
        {\empcmd{fill bpath jetset withcolor white;}%
         \empcmd{drawarrow (hconnect (events, .5, jetset, .5));}%
         \empcmd{drawboxed (jetset);}}
        {\empcmd{fill bpath whizard withcolor 0.9white;}%
         \empcmd{drawboxed (whizard);}}
        {\empcmd{drawarrow (vconnect (params, .5, whizard, .75));}%
         \empcmd{drawarrow (vconnect (cuts, .5, whizard, .70));}%
         \empcmd{drawarrow (vconnect (whizard, .4, sigma, .5));}%
         \empcmd{drawarrow (vconnect (whizard, .5, hist, .5));}%
         \empcmd{drawarrow (vconnect (whizard, .5, events, .5));}}%
        {\empcmd{fill bpath vamp withcolor 0.9white;}%
         \empcmd{drawboxed (vamp);}%
         \empcmd{drawdblarrow (hconnect (whizard, .5, vamp, .5));}}%
        {\empcmd{fill bpath circe withcolor 0.9white;}%
         \empcmd{drawboxed (circe);}%
         \empcmd{drawdblarrow (hconnect (circe, .5, whizard, .5));}}%
        {\empcmd{fill bpath omega withcolor 0.9white;}%
         \empcmd{drawboxed (omega);}%
         \empcmd{drawarrow (vconnect (feynman, .5, omega, .5));}%
         \empcmd{drawdblarrow (vconnect (omega, .5, whizard, .35));}}%
        {\empcmd{fill bpath madgraph withcolor white;}%
         \empcmd{drawboxed (madgraph);}%
         \empcmd{drawarrow (vconnect (feynman, .5, madgraph, .5));}%
         \empcmd{drawdblarrow (vconnect (madgraph, .5, whizard, .25));}}%
      \empcmd{setbounds currentpicture to (0,0)--(w,0)--(w,h)--(0,h)--cycle;}
    \end{empx}
  \end{center}
  \caption{\label{fig:whizard-components}%
    The event generator generator WHIZARD provides phase space
    parametrizations and utilizes external components for matrix
    element generation and adaptive multi channel Monte Carlo
    sampling.}
\end{figure}

WHIZARD~\cite{Kilian:WHIZARD} solves the other problem set forth in
the introduction: the efficient sampling of multi particle phase
space.  In addition, it provides a driver routine for the automated
construction of efficient unweighted event generators.  The unweighted
events are written either in ASCII or STDHEP~\cite{STDHEP} format.
Leading order initial state radiation, beamstrahlung (via
CIRCE~\cite{CIRCE}) and beam polarization are supported.

For a given process, WHIZARD automatically identifies the kinematical
variables in which singularities can appear.  In a second step,
WHIZARD constructs a set of phase space parameterizations in which all
singular variables appear explicitely.  In general, it is impossible
to find a single parameterization that includes all singular
variables~\cite{VAMP}, but it is always possible to cover them by a
finite set of parameterizations.  Finally, WHIZARD sets up the adaptive
multi channel sampling library \textit{VAMP}~\cite{VAMP} for this set
of parameterizations, calls an external matrix element generator and
creates an unweighted Monte Carlo event generator.  So far, this
approach has been shown to work well for processes with up to eight
particles in the final state.

Among matrix element generators, O'Mega is the preferred choice for
polarized scattering of many weakly interacting particles.  It
generates the most efficient code in this case and offers the greatest
flexibility for handling unstable vector bosons and for including some
deviations from the standard model.  But WHIZARD does not depend on
O'Mega and can use other matrix element generators as well. Indeed,
MADGRAPH~\cite{MADGRAPH:1994} is used for standard model amplitudes
with interfering color structures, while CompHEP~\cite{CompHEP} can be
more efficient for the scattering of few and unpolarized particles.

\section{Applications}
The first complete experimental study of vector boson scattering in
six fermion production for linear collider physics has been the first
serious application and is discussed elsewhere~\cite{WWto6f}.

The \emph{Higgsstrahlung} process $e^-e^+\to \nu_e\bar\nu_e b\bar b$
provides a simple example for a completely automated calculation
(cf.~Figure~\ref{fig:Higgsstrahlung}).  There
are 21~diagrams in four groves: {$5\times$}Higgs\-strahlung,
{$10\times$}$WW$-fusion, {$4\times$}$ZZ$ production,
{$2\times$}$Z$-bremsstrahlung.  These diagrams contribute
singularities in many time-like and space-like channels.
\begin{figure}
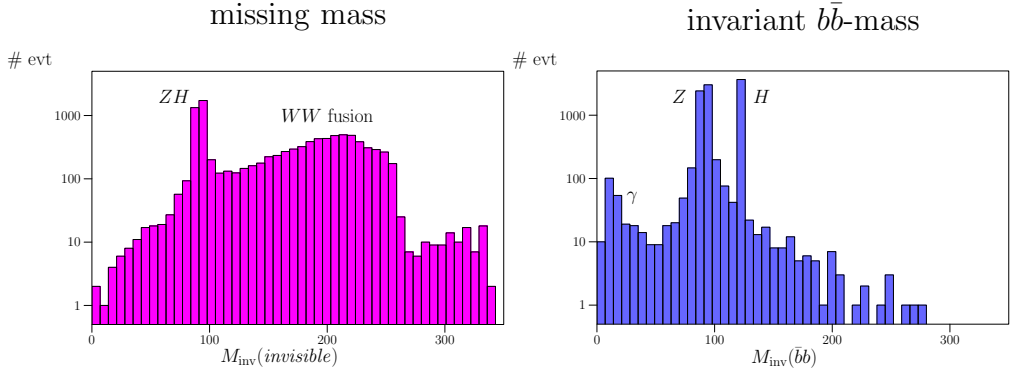

  \begin{center}
    \parbox{0.48\textwidth}{%
      \begin{center}
	missing mass\\[\baselineskip]
	\includegraphics[width=.4\textwidth]{nnbb-invis.eps}
      \end{center}}
    \parbox{0.48\textwidth}{%
      \begin{center}
	invariant $b\bar b$-mass\\[\baselineskip]
	\includegraphics[width=.4\textwidth]{nnbb-bb.eps}
      \end{center}}
  \end{center}
  \caption{\label{fig:Higgsstrahlung}
    Distributions of 10000 \emph{unweighted}
    $e^-e^+\to \nu_e\bar\nu_e b\bar b$
    events at~$\sqrt{s}=350\;\textrm{GeV}$
    for~$m_H=120\;\textrm{GeV}$,
    corresponding to~$16.68\,\textrm{fb}^{-1}$.}
\end{figure}

After adapting the grids with 20000 events with fixed relative weights
of the channels, an error in the total cross section
of~$2.2\%=3.76/\sqrt{N}$ is obtained with a projected efficiency
of~$2\%$ for unweighted event generation.  After ten steps of adapting
the relative weights, each using 20000 events, the relative error in
the total cross section for 20000 events is reduced
to~$0.48\%=0.68/\sqrt{N}$ and efficiency for unweighted event
generation is increased to~$19.3\%$.  This process consumes 24\,min
for adaptation and 5\,min for generating 10000 \emph{unweighted}
events on a Pentium~450\,MHz.  Since the adaption typically consumes
more than the subsequent event generation, the adapted grids and
weights can be saved and reloaded for generating sets of event samples
with similar parameters and cuts.

\subsection*{Acknowledgments}
I thank my O'Mega collaborators Mauro Moretti and J\"urgen Reuter.  I
thank Wolfgang Kilian for valuable suggestions and for ``early
adoption'' of O'Mega.  I am supported by the German Bundesministerium
f\"ur Bildung und Forschung (05\,HT9RDA) and Deutsche
Forschungsgemeinschaft (MA\,676/6-1).


\end{empfile}

\begin{thebibliography}{10}
  \bibitem{Ohl:2000:ACAT}
    T. Ohl, \textit{O'Mega: An~Optimizing Matrix~Element~Generator},
    Proceedings of the \textit{Workshop on Advanced Computing and
    Analysis Technics in Physics Research,} Fermilab, October 2000,
    IKDA 2000/30, hep-ph/0011243.
  \bibitem{O'Mega}
    M. Moretti, T. Ohl, and J. Reuter,
    (to be published),
    \url{http://heplix.ikp.physik.tu-darmstadt.de/~ohl/omega/}.
  \bibitem{Kilian:WHIZARD}
    W. Kilian, (to be published),
    \url{http://www-ttp.physik.uni-karlsruhe.de/~kilian/whizard/}.
  \bibitem{VAMP}
    T. Ohl, Comput.\ Phys.\ Commun.\ \textbf{120} (1999) 13.
  \bibitem{ALPHA:1997/HELAC:2000}
    F. Caravaglios and M. Moretti, Z.{} Phys.{} \textbf{C74} (1997) 291;
    A. Kanaki and C. Papadopoulos, DEMO-HEP-2000/01, hep-ph/0002082,
    February 2000.
  \bibitem{MADGRAPH:1994}
    T. Stelzer and W.F. Long,
    Comput.{} Phys.{} Commun.{} \textbf{81} (1994) 357.     
  \bibitem{STDHEP}
    L. Garren, \url{http://www-pat.fnal.gov/stdhep.html}.
  \bibitem{CIRCE}
    T. Ohl, Comput.\ Phys.\ Commun.\ \textbf{101} (1997) 269.
  \bibitem{CompHEP}
    E. E. Boos et al, \textit{CompHEP - a package for evaluation of
    Feynman diagrams and integration over multi-particle phase space,}
    hep-ph/9908288.
  \bibitem{WWto6f}
    R. Chierici, these proceedings; R. Chierici and S. Rosati, (to be published).
\end{thebibliography}
\end{document}